\documentclass[twocolumn]{aastex631}
\pdfoutput=1 
\usepackage{appendix}
\usepackage{natbib}
\usepackage{graphicx}
\usepackage{enumitem}
\usepackage{hyperref}
\usepackage{mathtools, amsmath,amstext}
\usepackage{cleveref}
\usepackage[T1]{fontenc}
\usepackage[figure,figure*]{hypcap}

\newcommand{\G}{\mathcal{G}}

\newcommand{\appropto}{\mathrel{\vcenter{\offinterlineskip\halign{\hfil$##$\cr\propto\cr\noalign{\kern2pt}\sim\cr\noalign{\kern-2pt}}}}}

\graphicspath{{./}{figures/}}

\accepted{May 31, 2022}
\submitjournal{ApJ}
\shorttitle{Generalized Peas-in-a-Pod}
\shortauthors{Goyal \& Wang}

\begin{document}

\title{Generalized Peas-in-a-Pod: Extending Intra-System Mass Uniformity to Non-TTV Systems via the Gini Index}

\author[0000-0001-9652-8384]{Armaan V. Goyal}
\affil{Department of Astronomy, Indiana University, Bloomington, IN 47405}

\author[0000-0002-7846-6981]{Songhu Wang}
\affil{Department of Astronomy, Indiana University, Bloomington, IN 47405}




\begin{abstract}
\noindent 

It has been demonstrated that planets belonging to the same close-in, compact multiple-planet system tend to exhibit a striking degree of uniformity in their sizes. A similar trend has also been found to hold for the masses of such planets, but considerations of such intra-system mass uniformity have generally been limited to statistical samples wherein a majority of systems have constituent planetary mass measurements obtained via analysis of transit timing variations (TTVs). Since systems with strong TTV signals typically lie in or near mean motion resonance, it remains to be seen whether intra-system mass uniformity is still readily emergent for non-resonant systems with non-TTV mass provenance. We thus present in this work a mass uniformity analysis of 17 non-TTV systems with masses measured via radial velocity (RV) measurements. Using the Gini index, a common statistic for economic inequality, as our primary metric for uniformity, we find that our sample of 17 non-TTV systems displays intra-system mass uniformity at a level of $\sim 2.5 \sigma$ confidence. We provide additional discussion of possible statistical and astrophysical underpinnings for this result. We also demonstrate the existence of a correlation ($r = 0.25$) between characteristic solid surface density ($\Sigma_0$) of the minimum mass extrasolar nebula (MMEN) and system mass Gini index, suggesting that more massive disks may generally form systems with more unequal planetary masses.

\end{abstract}



\section{Introduction} \label{sec:intro}

The architectural characterization of extrasolar planet systems is a key component of understanding their formation histories, and the study of trends in such architectures emergent over a large sample of systems serves to place powerful constraints on their possible assembly mechanisms and evolutionary pathways. The mission data from NASA's \textit{Kepler} Space Telescope \citep{borucki}, which comprises the largest and most comprehensive exoplanet catalog to date, displays an abundance of systems with multiple close-in ($P \lesssim 100$ days), tightly-spaced planets with sizes between that of Earth and Neptune ($1 R_{\oplus} \lesssim R_{p} \lesssim4 R_{\oplus}$) on nearly-coplanar, almost-circular orbits (\citeauthor{xie} \citeyear{xie}; \citeauthor{thompson} \citeyear{thompson}; \citeauthor{weiss_rad} \citeyear{weiss_rad}; \citeauthor{millholland2021} \citeyear{millholland2021}). Such systems with multiple super-Earths or sub-Neptunes thus appear to be the most common type of planetary systems in the galaxy, and it has been shown that such systems tend to exhibit a surprising degree of uniformity in their constituent planet sizes \citep{weiss_rad}. This assessment of size similarity was complemented by \citet{millholland}, who demonstrated an analogous uniformity for planetary masses that are determined via transit timing variations (TTVs).

The striking prevalence of this "peas-in-a-pod"-style uniformity of planets within \textit{Kepler} multiple-planet systems has led to various series of inquiry probing the nature of the trend itself. Recent statistical studies have confirmed that the observed preferences for size uniformity may be recovered via full forward modeling and considerations of complexity theory (\citeauthor{he} \citeyear{he}, \citeauthor{gilbert} \citeyear{gilbert}), while a theoretical framework built upon energy optimization and total system mass budgeting also serves to reproduce and physically motivate the peas-in-a-pod configuration \citep{adams}. \citet{millholland_split} have also explored the existence of a "split peas-in-a-pod" phenomenon within such multiple-planet systems, noting that the distinct sub-groupings of rocky super-Earths and gaseous sub-Neptunes within a given system each exhibit a greater degree of size similarity than the system as a whole.

While it is clear that the peas-in-a-pod trend is generally well-characterized in many regards, it should be noted that analyses of intra-system mass uniformity have largely been limited by the exclusive consideration of planets with TTV-derived masses (hereafter TTV systems). Since systems with strong TTV signals typically indicate an orbital configuration near mean motion resonance (MMR), it remains to be seen whether intra-system mass uniformity is still readily emergent for non-resonant systems with non-TTV mass provenance (hereafter non-TTV systems) that exhibit less overall proximity to MMR. Such a consideration is further motivated by the fact that our own solar system, which displays considerable global diversity in planetary mass, also exhibits a high degree of mass uniformity for neighboring objects in or near MMR, such as the Venus-Earth (period ratio near 13:8) and Uranus-Neptune (near 2:1) systems, as well as the Galilean moons Io, Europa, and Ganymede (near 4:2:1). It is therefore necessary for the generalization of the peas-in-a-pod effect to observe if mass uniformity is unique to systems near MMR.

We note that a recent study by \citet{otegi} explored such a notion of generalized intra-system mass uniformity via added consideration of planets with radial velocity (RV) mass measurement. Although such an inclusion is invaluable in its allowance for the presence of non-TTV systems in a given sample, any existing relationships between mass uniformity and MMR may only be probed implicitly, as mass uniformity for a mixed sample containing both TTV and non-TTV systems may simply be an artifact of the former as opposed to a specifically-distinguishable property of the latter. Since 27 of the 48 systems included in \citet{otegi} display measurable TTVs, their demonstration of intra-system mass uniformity provides strong evidence for its generalization beyond TTV systems alone, but does not disentangle the TTV and non-TTV populations to draw specific physical conclusions regarding resonance. As such, the primary concern of our analysis is to probe intra-system mass uniformity for \textit{exclusively} non-TTV systems, an astrophysically-distinct population for which such uniformity has not yet been uniquely assessed.

We thus present in this work a direct assessment of intra-system mass uniformity for 17 strictly non-TTV systems, or transit systems that do not display any significant TTVs with precise mass measurements obtained by the RV technique. Using the Gini index, a commonly used statistic for measuring income inequality in the field of econometrics, as our primary metric for uniformity, we find that our sample displays intra-system mass uniformity with $\sim 2.5\sigma$ confidence when compared to a null hypothesis of random assortment. We provide a general motivation for our use of the Gini index (Section \ref{sec:gini}) and a brief overview of our sample selection (Section \ref{sec:sample}) before the presentation and discussion of our results (Sections \ref{sec:analysis} and \ref{sec:discussion}).
    
\section{Gini Index} \label{sec:gini}
Throughout the entirety of our analysis, we adopt the Gini index as our sole metric for assessing intra-system uniformity. The index, introduced by \citet{gini}, is a statistic commonly used in econometrics to quantify the degree of income or wealth inequality present in a given population. The mathematical statement of the Gini index for a data vector $x$ of size $N$ can be expressed as:

\begin{equation} \label{eq:1}
    G = \frac{1}{2N^{2}\overline{x}}\sum_{i=1}^{N} \sum_{j=1}^{N} |x_{i}-x_{j}|.
\end{equation}
The expression above is equivalent to \textit{half of the relative mean absolute difference} of the sample $x$, or put more simply, half of the average difference between all possible pairwise combinations of the data, normalized to the mean $\overline{x}$. The index is intrinsically normalized to unity, such that a completely uniform dataset would yield $G = 0$ and a maximally unequal dataset would yield $G = 1$. 

Despite the simplicity of the metric, the Gini index is known to be a downward-biased statistic, in that the true degree of inequality for a given population will consistently be underestimated for small sample sizes. \citet{deltas_2003} demonstrates this bias via Monte Carlo simulation: by calculating the average Gini index for progressively larger samples drawn from each of a uniform, log-normal, and exponential distribution, it was found that the true (asymptotic) Gini of the population can generally be estimated with $\leq 1\%$ accuracy for $N \gtrsim 50$, and that the index will typically serve to underestimate the true Gini by $\geq 15\%$ for $N \lesssim 10$. Fortunately, \citet{deltas_2003} also provides a simple renormalization to mitigate the effects of this small-sample bias, defining an adjusted Gini index:

\begin{equation} \label{eq:2}
    \mathcal{G} = \left(\frac{N}{N-1}\right)G.
\end{equation}

Since our analysis applies considerations of the Gini index to individual planet systems, the vast majority of which host a number of planets $N_{p} \lesssim 5$, we note that use of this correction term will likely be imperative. 

\begin{figure}
    \centering
    \includegraphics[width=0.45\textwidth]{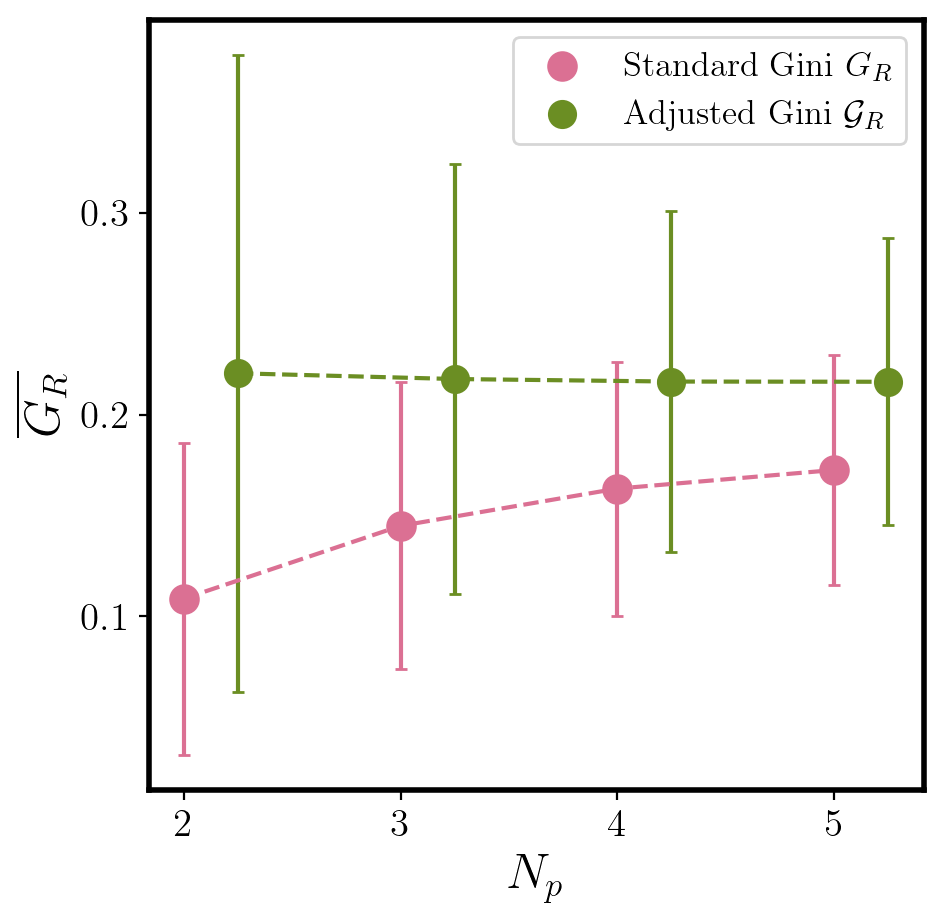}
    \caption{The standard Gini index will heavily underestimate the true degree of non-uniformity in systems containing $N_{p} \lesssim 5$ planets, while the adjusted Gini index exhibits no such bias for the same systems. Each pink point represents the mean standard Gini $G_{R}$ of 1000 random sets of $N_{p}$ radii drawn from a sample of CKS multiple-planet systems. We plot green points the adjusted Gini index $\mathcal{G}_{R}$ that includes the correction term present in (\ref{eq:2}). The standard Gini displays variation of $\sim 37\%$ over $N_{p} \in [2, 5]$ while the  adjusted Gini successfully is stabilized to within $\sim 2 \%$.}
    \label{fig:1}
\end{figure}

To determine the precise impact of this bias and the efficacy of the correction term, we perform a similar Monte Carlo process on the planetary radius data for a collection of multiple-planet systems from the California-Kepler Survey (CKS) sample presented in \citet{weiss_samp}. For each system planet multiplicity $N_{p} \in [2, 5]$, we generate $10^{4}$ mock systems of $N_{p}$ planets with corresponding values of $R_{p}$ drawn randomly from the CKS population. We calculate the average Gini and adjusted Gini across the $10^{4}$ mock samples for each value of $N_{p}$; the results are plotted in Figure \ref{fig:1}. We see that compared to the null expectation of no variation in the Gini index for $N_{p} \in [2, 5]$, the standard Gini index $G$ indeed maintains its bias within the given CKS population, but that the adjusted index $\mathcal{G}$ effectively stabilizes (within $\sim 2\%)$ the metric even at the smallest possible values of $N_{p}$. We shall therefore use adjusted index $\mathcal{G}$ for all considerations of the Gini index within our analysis.

\section{Sample Selection} \label{sec:sample}
For the purposes of our analysis, we aim to construct a sample of systems that abide by the following criteria:
\begin{enumerate}
    \item All planets are transiting.
    \item All planets have RV mass measurements with fractional error unity or less.
    \item No planets show TTVs (adopted observational indicator for systems that are not near MMR).
    \item System is compact; all neighboring period ratios are less than $6$ (see \citeauthor{wang_148} \citeyear{wang_148} for the definition of compact multi-planet system). 
\end{enumerate}

While the motivations for first three criteria have since been discussed, we impose our compactness cut as a means of specifically selecting close-in systems whose planets likely share a common and generally quiescent formation history. Possible long-period companions to otherwise compact multiple-planet systems may have experienced more isolated formation in a different region of the protoplanetary disk. Most commonly, such long-period planets are often gas giants that form beyond the snowline of their system, and therefore represent an entirely different regime of both size and composition when compared to their close-in companions. For systems that instead contain ultra-short period (USP) members, such as TOI-561 (\citealt{weiss_561}, \citealt{lacedelli}), it has been shown that the USP planets are often markedly smaller than other members of the same system, and may have experienced a much more dynamically volatile formation history. \citep{winn_usp}.

We query the NASA Exoplanet Archive\footnote{\label{note1}\url{exoplanetarchive.ipac.caltech.edu/docs/data.html}} for our initial sample and subsequently impose the aforementioned error, TTV, and compactness restrictions. We shall note here that as a result of our sample being drawn from the NASA Exoplanet Archive, the corresponding planetary data represent an amalgamation of values derived from different surveys, instruments, and analysis techniques. Any population study performed on such a heterogenous sample would therefore be subject to complex biases resulting from inherent differences in systematics and heuristic determination of planetary parameters, as well as non-standardized determination of errors across the entire sample. The provision of a homogeneous RV exoplanet sample, such as that presented by the California Legacy Survey \citep{rosenthal}, is therefore extremely valuable in its minimization of these biases and the resulting reduction of statistical noise in an associated population analysis. While further exploration of the biases generated by sample heterogeneity and their potential effects are beyond the scope of this work, we nonetheless acknowledge the presence of this caveat as it pertains to our own sample and analysis.

\begin{figure}
    \centering
    \includegraphics[width=0.5\textwidth]{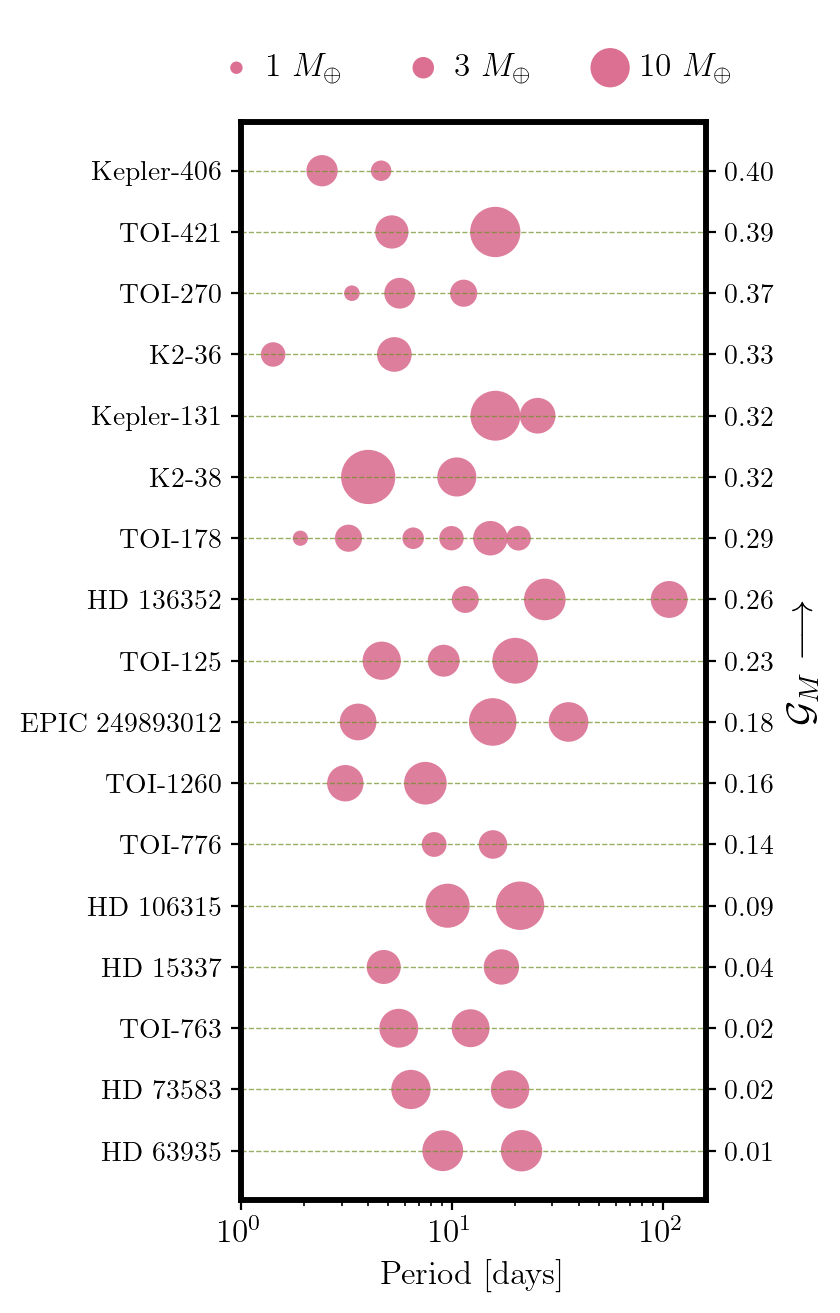}
    \caption{Visual representation of intra-system mass uniformity for non-TTV systems with RV-determined masses via plotted orbital architectures of the 17 systems in our primary sample. The point sizes correspond to  planetary mass $M_{p}$. Reference masses of $1 M_{\oplus}$, $3 M_{\oplus}$, and $10 M_{\oplus}$ are included for scale. The systems are plotted in order of their adjusted mass Gini $\mathcal{G}_{M}$, from most diverse to most uniform. We note the presence of intra-system mass uniformity at the 2.5 $\sigma$ confidence level for these 17 systems (Figure \ref{fig:3} and Section \ref{sec:analysis}).}
    \label{fig:2}
\end{figure}

While not explicitly mentioned in our selection criteria, we also omit incomplete systems for which not all constituent planets were present in our initial query. If a system contains incomplete data such that at least one constituent planet lacks a well-defined mass measurement, then its inclusion in the analysis will introduce a bias that serves to artificially enhance the significance of intra-system mass uniformity for the total sample. This bias is best understood as a selection effect which posits that the lack of a true mass measurement for a given planet may occur as a result of discrepancy from other members of the same system in terms of its orbital or dynamical characteristics, prominent examples of such being unusually small planets, as in the case of USPs, and inclined, non-transiting planets, which are generally of higher mass. The regular exclusion of entire systems containing such planets may indeed introduce its own effects via their under-representation in the sample compared to the true planetary population at large. However, omission of the individually-discrepant planets alone with continued use of their system companions introduces a more directly-pronounced bias with regard to intra-system mass uniformity, as the removal of a given discrepant planetary mass will cause the remaining, incomplete system to be biased towards a greater degree of uniformity than is physically accurate.

The resulting sample contains 15 close-in, compact, non-TTV systems with 38 total planets.  We note that our initial query is agnostic to any newly-confirmed systems discovered by the Transiting Exoplanet Survey Satellite (TESS) that are not yet present in the NASA Exoplanet Archive. HD 73583 \citep{barr} and HD 63935 \citep{scarsdale} are two such non-TTV systems that similarly abide by our restrictions and are therefore also considered for our analysis. The resulting final sample thus consists of 17 close-in, compact, non-TTV systems containing a total of 42 planets. Figure \ref{fig:2} illustrates the orbital configurations and adjusted mass Gini indices of these 17 systems, with point sizes corresponding to planetary mass.

For the sake of providing more intuitive context for both the mass uniformity of these multiple-planet systems and the Gini index in general, we will note here the adjusted mass Gini indices for some groupings of solar system objects. The inner four solar system planets ($\mathcal{G}_{M} = 0.60$) illustrate a non-resonant example while the Earth-Venus ($\mathcal{G}_{M} = 0.10$) and Uranus-Neptune ($\mathcal{G}_{M} = 0.08$) pairs, as well as the four Galilean moons ($\mathcal{G}_{M} = 0.27$), exemplify systems in or near MMR. We note that each of the resonant systems fall within the range displayed by the systems in Figure \ref{fig:2}, while the inner solar system exhibits a greater degree of mass diversity than any system contained in our sample.

\section{Analysis and Results} \label{sec:analysis}
Our primary null hypothesis test was constructed similarly to the analysis presented in \citet{millholland}: for the set of 17 close-in, compact, non-TTV systems and 42 total planets in our sample, we performed $10^{5}$ control realizations wherein the total number of systems and the multiplicity of each system was preserved, but the $M_{p}$ values of the specific planets within them would be chosen at random from the aggregate population without replacement. The adjusted mass Gini indices ($\mathcal{G}_{M}$) were then calculated for and summed across all 17 mock systems within each realization to yield a final control distribution, illustrated by the pink histogram in Figure~\ref{fig:2}; the $\mathcal{G}_{M}$ summation was then performed for the 17 real, non-shuffled systems in our sample for comparison with the controls, and is represented by the solid black line. We see that the total mass Gini across the 17 real systems $(\sum \mathcal{G}_{M} = 3.59)$ falls well below the bulk of the distribution generated from our $10^{5}$ control shufflings $(\sum \mathcal{G}_{M} = 5.00 \pm 0.56)$, and that the null hypothesis of random intra-system mass assortment is therefore rejected with $\sim 2.5 \sigma$ confidence. 

\begin{figure}
    \centering
    \includegraphics[scale=0.68]{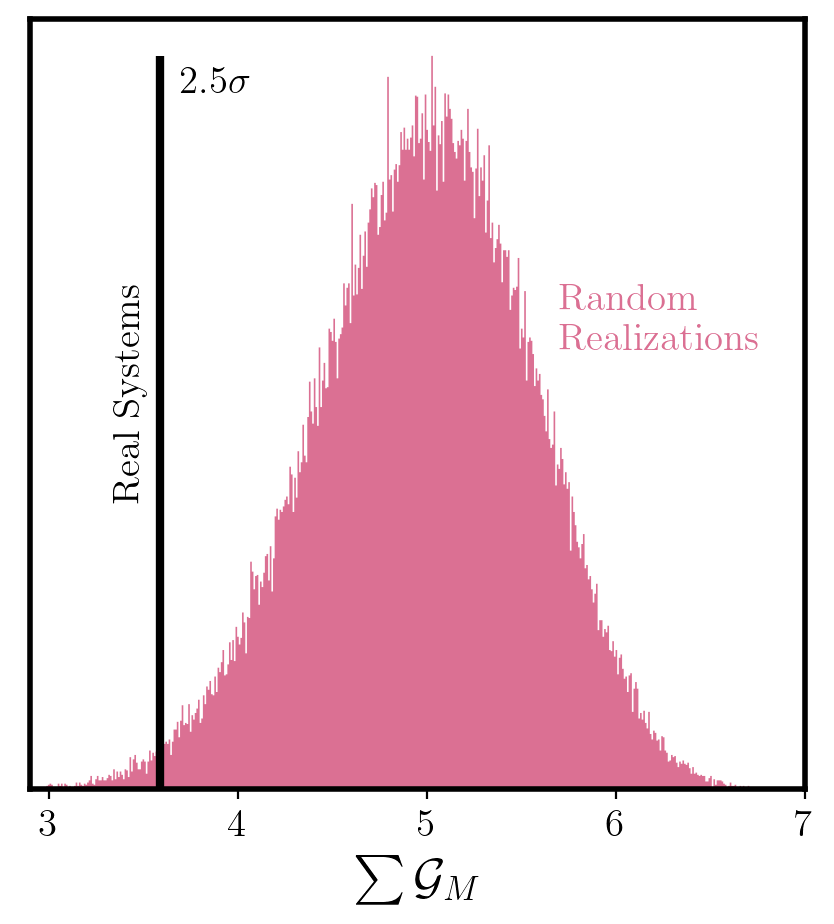}
    \caption{Statistical validation of intra-system mass uniformity for $17$ non-TTV systems at $\sim 2.5 \sigma$ confidence compared to random expectation. Pink histogram illustrates control distribution of the total adjusted mass Gini $\sum \mathcal{G}_{M}$ for $10^{5}$ random shufflings of the planets in our 17 non-TTV systems. The control realizations preserve the overall number of systems $N_{sys}$ and the number of planets $N_{p}$ within each system, but the $M_{p}$ values for each planet are drawn at random (without replacement) from the total planet population in our sample. The solid black line represents $\sum \mathcal{G}_{M}$ for our 17 real non-TTV systems, which we falls well below the bulk of the control distribution to signify a greater degree of uniformity.}
    \label{fig:3}
\end{figure}

As a means of ruling out detection bias as the most prominent cause for the emergence of this intra-system mass uniformity, we perform an additional null hypothesis test that considers only mock systems wherein all constituent planets have a RV signal greater than $1\,{\rm m\,s^{-1}}$, the minimum value observed across the 42 real planets in our sample. Following \citet{weiss_pet}, we note that shuffling of discrete planetary data for such a detection-limited null hypothesis test may result in artificially-suppressed representation of small $\bm{(R < R_{\oplus})}$ planets in the mock systems, so we instead draw planetary radii randomly from a log-normal distribution $\bm{\ln(R_{p}/R_{\oplus}) \in \mathcal{N}(\mu, \sigma^{2})}$ with $\bm{\mu = 0}$ and $\bm{\sigma = 1}$. Planetary masses are then determined probabilistically via a mass-radius relation \citep{chen_kipping}. For any mock planetary mass that yields RV signal below our established minimum of $1\,{\rm m\,s^{-1}}$, the radius is redrawn until the detection criterion is satisfied. We construct in this manner 1000 detection-limited mock samples, once again preserving from our real sample (Figure \ref{fig:2}) the total number of systems and planet multiplicities therein. We then replicate our null hypothesis test (Figure \ref{fig:3}) by calculating the total mass Gini for each of these detection-limited mock samples, for which we obtain a control distribution of $\sum \mathcal{G}_{M} = 5.00 \pm 0.70$. Our real sample ($\sum \mathcal{G}_{M} = 3.59$) thus rejects the detection-limited null hypothesis with $\sim 2 \sigma$ confidence, indicating that the degree of intra-system mass uniformity displayed by our 17 real non-TTV systems holds a $\lesssim 5$\% probability of being reproduced by detection bias alone.

\section{Discussion} \label{sec:discussion}
The primary implication of our result may be expressed most simply as the following: while it is known that intra-system mass uniformity is readily observed for systems in or near MMR with TTV-derived masses, \textit{we find that this uniformity may be maintained for close-in, compact, non-TTV systems with RV-based mass measurements as well}. 

\newpage 

\subsection{Statistical and Observational Biases} \label{subsec:statobs}

\begin{figure}
    \centering
    \includegraphics[scale=0.7]{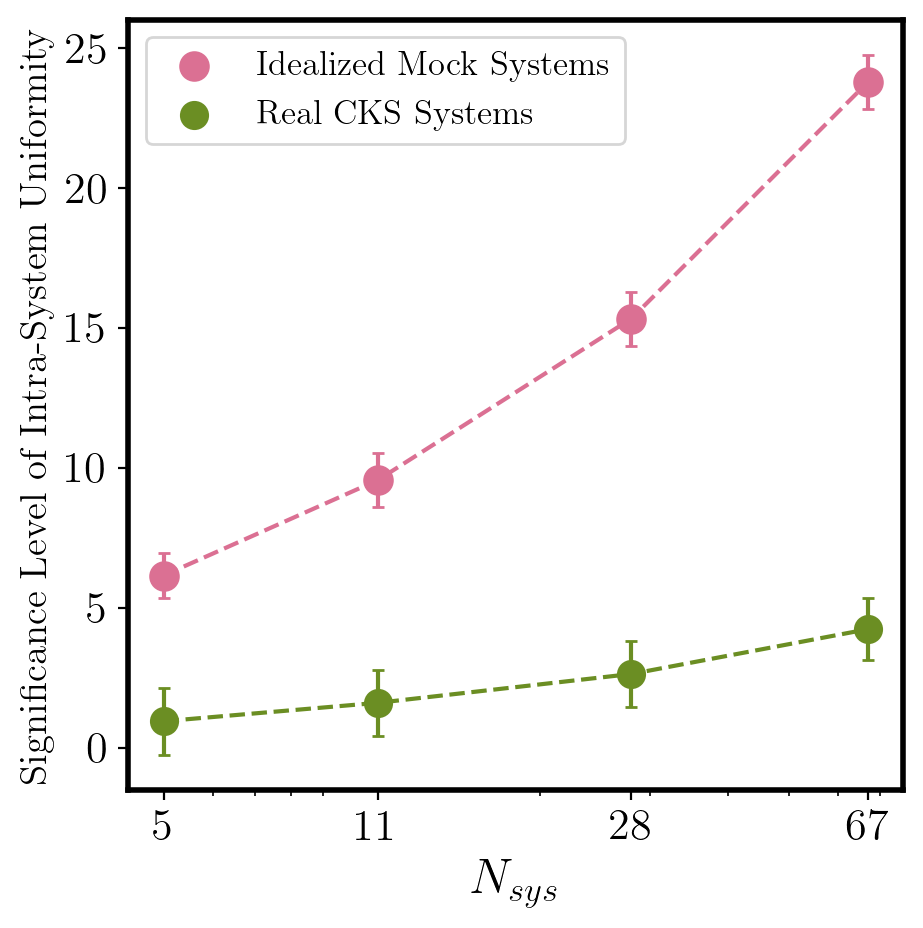}
    \caption{Total significance of intra-system uniformity is biased towards larger values for samples containing more systems (larger $N_{sys}$). Tested on planetary radius data for 299 CKS systems from Section \ref{sec:gini} (green points), and a mock population of 300 "ideal" peas-in-a-pod systems (pink points). At each value of $N_{sys}$, we perform 1000 random draws of $N_{sys}$ systems from the relevant population and subject these systems to the null hypothesis testing described in Figure \ref{fig:3} and Section \ref{sec:analysis}. We see a monotonic rise in total uniformity significance with increasing $N_{sys}$ for both populations.}
    \label{fig:4}
\end{figure}

We note that the $\sim 2.5 \sigma$ result for our 17 close-in, compact, non-TTV systems sits below the $\sim 7.6 \sigma$ figure determined by \citet{millholland} for a sample of 37 TTV systems. In exploring whether the relation between these results is statistical or astrophysical in nature, we posit that the difference in significance levels may be caused by the difference in sample size between the respective system populations. In order to test for the existence of such sample-size effects, we refer again to the precise planetary radius dataset corresponding to the CKS systems described in Section \ref{sec:gini}. We fix trial system sizes of $N_{sys} \in [5, 11, 28, 67]$ to provide  values that are equally distributed in logarithmic space between 0 and 299, where the latter is the total number of systems present in the CKS dataset . We perform 1000 random draws of $N_{sys}$ systems from the CKS population and subject each set of $N_{sys}$ systems to the same null hypothesis testing described in Section \ref{sec:analysis}, thus generating a distribution of significance levels at each $N_{sys}$. We plot these distributions in Figure \ref{fig:4}, where we see that the mean significance indeed increases with rising number of systems $N_{sys}$.

To determine whether this effect is itself a physical artifact of the \textit{Kepler} sample or purely statistical in nature, we repeat this testing for a mock sample of "ideal" peas-in-a-pod systems. We generate 300 mock three-planet systems each with a single fixed planet radius $\tilde{R_{p}} \in [1R_{\oplus}, 4R_{\oplus}]$, such that the 300 systems are spaced with $\Delta \tilde{R_{p}} = 0.01 R_{\oplus}$. We add Gaussian scatter to each individual planet at the level of $\delta R_{p} = \Delta \tilde{R_{p}}/10 = 0.001 R_{\oplus}$ such that the set of planet radii $\{R_{p}\}$ within each system are not exactly equal but still exhibit far less variation than that of the prescribed $\tilde{R_{p}}$ between systems. We repeat the generation of our significance distributions for this idealized sample and plot the results in Figure \ref{fig:4}. We observe that while the significance values for the idealized sample are systematically higher as expected, the overall trend of significance rising with $N_{sys}$ is preserved. We therefore state that our $\sim 2.5 \sigma$ result is influenced, at least in part, by our limited sample size.

We may probe such notions more directly by comparing the degree of intra-system mass uniformity for our non-TTV systems to that of random subsets of equivalent size drawn from the \citet{millholland2021} TTV sample. From the 37 total systems in \citet{millholland}, we perform 1000 random draws of 17 systems from and calculate the total mass Gini for each subsample. We obtain a distribution of $\sum \mathcal{G}_{M} = 4.72 \pm 0.70$ for these TTV subsamples, which, upon recalling the analogous figure for the 17 non-TTV systems considered in this work ($\sum \mathcal{G}_{M} = 3.59$), suggests that the respective degrees of intra-system mass uniformity for our non-TTV sample and a similarly-sized TTV sample are indeed fairly comparable.

Nevertheless, we cannot rule out the possibility that the difference between our non-TTV result and that of the \citet{millholland} TTV sample may be the result of additional observational biases inherent to either system type. \cite{steffen} illustrates via Monte Carlo simulation that while both TTV and RV methods are themselves robust estimators of planetary mass, RV analysis is systematically biased towards the detection of more massive or denser planets. We demonstrate in Section \ref{subsec:disk} that systems with a greater total planetary mass may have more unequal constituent planet masses, suggesting that this high-mass bias of RV-dominated systems can generate an aggregate bias towards lower significance values for total intra-system mass uniformity in a given population. 

\newpage 

\subsection{Energy Optimization}
\label{subsec:energy}
The prevalence of intra-system mass uniformity for our close-in, compact, non-TTV systems provides considerable agreement with the theoretical framework presented in \citet{adams}, who suggested that the nearly equal-mass planets characteristic of the peas-in-a-pod architecture may constitute a minimum-energy optimum state for nascent planets whose achievement is governed primarily by the total system planetary mass budget. In the case of coplanar, circular orbits, a given two-planet system with total planet mass budget $M_{T} \equiv M_{1} + M_{2}$, fixed spacing parameter $\Lambda \equiv a_{2}/a_{1}$, mass fraction $f \equiv M_{1}/M_{T}$, and approximate size equality $R_{1} \approx R_{2} \approx R_{p}$ will have a pair of dichotomous optima for its total energy (\citealt{adams}, \citealt{adams2}): a case of orbital energy domination for which the optimal mass fraction $f_{0}$ can be expressed as:

\begin{equation}
    f_{0}(\Lambda) = \frac{\Lambda + \sqrt{\Lambda} - 2}{3(\Lambda - 1)},
    \label{eq:4}
\end{equation}
and a case in which the binding energy dominates instead and $f_{0} \rightarrow 0$. \citet{adams} demonstrates that the transition between these two regimes occurs at a critical mass threshold for $M_{T}$ given by:

\begin{equation}
M_{C} = M_{*}\left( \frac{R_{p}}{a_{2}} \right) \frac{(\sqrt{\Lambda} + 2)(\sqrt{\Lambda}-1)^{2}}{4\alpha_{g}\sqrt{\Lambda}},
\end{equation}
which for typical values of $\Lambda = 1.6$, $M_{*} = M_{\odot}$, $a_{2} = 0.1$ AU, $R_{p} = 3R_{\oplus}$, and $\alpha_{g} = 0.5$ they find $m_{C} \approx 40 M_{\oplus}$.

For the case of a mass budget $M_{T} \gtrsim 40 M_{\oplus}$, we see that the optimal state of $f_{0} \rightarrow 0$ corresponds to a configuration in which $M_{2} \gg M_{1}$, and a single planet dominates the total mass budget to yield a maximally non-uniform system, similar to what was demonstrated by \cite{wang}. In the $M_{T} \lesssim 40 M_{\oplus}$ case, which is the typical mass budget regime for a given pair of Kepler planets, we see that the expression in Equation~(\ref{eq:4}) will approach $f_{0} \sim 1/2$, or $M_{1} \approx M_{2}$, for typical \textit{Kepler} values of $1.2 \lesssim \Lambda \lesssim 1.8$ \citep{weiss_rad}, confirming that the expectation for a pair of neighboring \textit{Kepler} planets would indeed be masses of approximately equal value. 

We immediately see that this energy optimization formalism predicts intra-system mass uniformity based on a simple threshold that is itself primarily a function of total system mass, and therefore invokes no considerations of MMR or associated formation dynamics. For these notions to manifest observationally, one would expect that systems with planet pairs above this mass threshold would contain giant planets that disrupt mass uniformity, while those largely beneath the mass threshold would be expected to exhibit mass uniformity regardless of resonant construction or the presence of TTVs. The first case has since been confirmed by \citet{wang}, who demonstrated that the presence of one or more Jovian-mass ($M > 100 M_{\oplus}$) planets in a system causes a breakdown of intra-system mass uniformity. The second case is confirmed in Section 5.1 of this work, where we verify that intra-system uniformity is present to a nearly identical degree within equally-sized samples of TTV and non-TTV systems that do not contain gas giants. Based on the emergent evidence for both predictions of this energy optimization model, we suggest that \textit{the presence of intra-system mass uniformity for compact, close-in, systems with multiple super-Earths or sub-Neptunes may be primarily governed by the total planetary mass budget of the system itself, regardless of resonance.}

\subsection{Mass Uniformity and Morphology of the Protoplanetary Disk}
\label{subsec:disk}
A possible corollary of the ramifications of the total planetary mass budgeting presented in \citet{adams} is the dependence of intra-system mass uniformity on the surface density profile of the protoplanetary disk. We attempt to probe this notion by modeling the minimum mass extrasolar nebula (MMEN) for a large sample of multi-planet systems and assessing any emergent correlations of the characteristic solid surface density ($\Sigma_{0}$) and MMEN slope ($\alpha$) with system mass Gini index. Since construction of the MMEN is largely agnostic to considerations of MMR, we allow for the inclusion of TTV systems into the relevant sample. However, in order to remain in accord with the regime of interest for considerations of mass budgeting, we impose an additional constraint of $M_{p} \leq 40 M_{\oplus}$. Our adjusted criteria yield for a total of 48 systems to be considered for our MMEN analysis.

We follow the construction of the MMEN put forth by \citet{chiang}, where, in a given system, we assign to each planet a surface density of

\begin{equation}
    \Sigma_{i} = \frac{M_{p,i}}{2 \pi a_{i} \Delta a_{i}} \approx \frac{M_{p,i}}{2 \pi a_{i}^{2}},
\end{equation}
where $M_{p,i}$ is the planetary mass and $a_{i}$ is the orbital semi-major axis. Note that we similarly adopt from \citet{chiang} the assumptions that close-in super-Earths are generally of near-solar composition and are primarily chondritic such that $\Sigma_{i}$ is mostly representative of solid mass, and that taking $\Delta a_{i} \approx a_{i}$ does not appreciably alter the distribution of calculated surface densities (additionally demonstrated in \citealt{dai}).

We fit the resulting solid surface density profiles ($\Sigma$ vs. $a$) for each system with a simple power law:

\begin{equation}
    \log(\Sigma) = \alpha \log(a) + \log(\Sigma_{0}).
\label{eq:8}
\end{equation}
Assuming $\delta \Sigma_{i}/\Sigma_{i} = \delta M_{p,i}/M_{p,i}$, we randomly sample the $\Sigma_{i}$ uniformly from $\Sigma_{i} \pm \delta \Sigma_{i}$ and obtain model parameters $\alpha$ and $\log(\Sigma_{0})$ from the power-law fit; we perform 1000 iterations of this two-parameter fitting process to generate for each system model parameter distributions $\alpha \pm \delta \alpha$ and $\log(\Sigma_{0}) \pm \delta \log(\Sigma_{0})$, where the uncertainties are representative of $1\sigma$ standard deviations. 

\begin{figure}[h]
    \centering
    \includegraphics[scale=0.7]{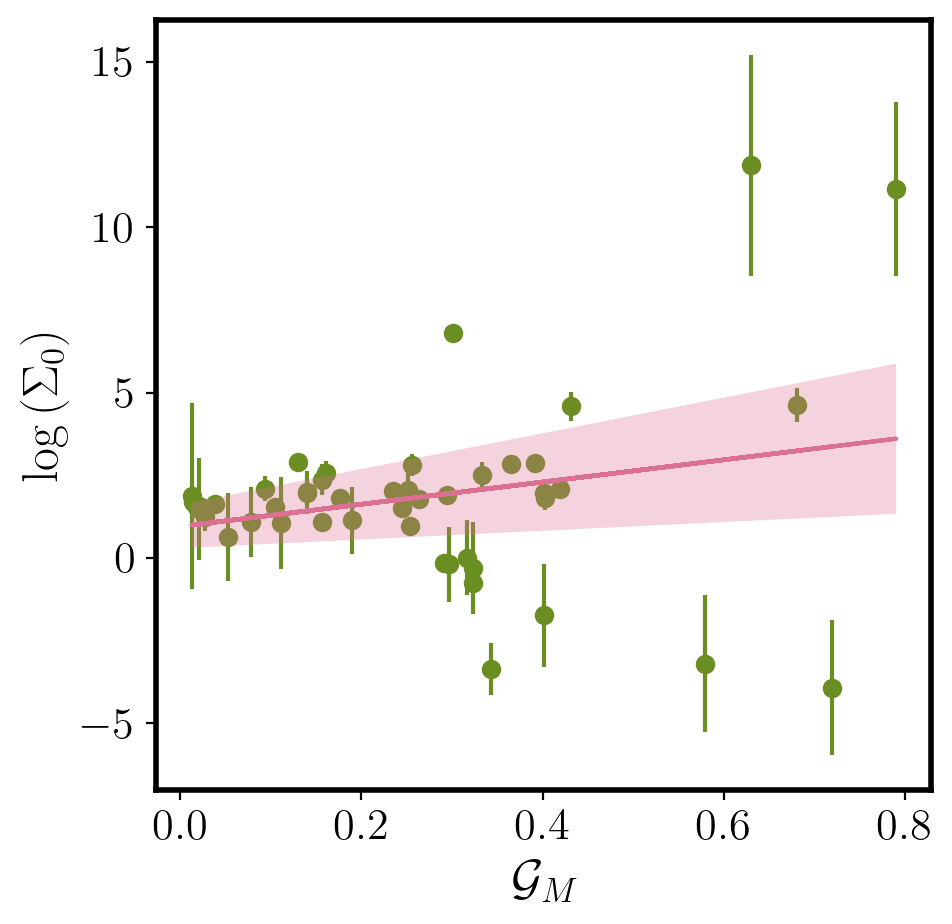}
    \caption{Characteristic solid surface density $\log(\Sigma_{0})$ of the MMEN displays a positive correlation ($r=0.25$) with adjusted mass Gini $\mathcal{G}_{M}$, indicating that more massive disk may result in the formation of systems with more unequal planet masses. The green points represent the 48 systems in our sample, and pink line and shaded region represent the best linear fit to the data and associated fit error. $1\sigma$ error bars in $\log(\Sigma_{0})$ are computed for each system via bootstrapping with 1000 fitting iterations. The $1\sigma$ error band on the $\log(\Sigma_{0})$-$\mathcal{G}_{M}$ linear fit is determined via similar bootstrapping methods.}
    \label{fig:5}
\end{figure}

Concerning the value of the MMEN slope $\bm{\alpha}$, we note that a system with perfectly-uniform planetary masses would have $\bm{\alpha = -2}$, while the minimum-mass solar nebula (MMSN) and the mean \textit{Kepler} MMEN (averaged over all planets) are both generally consistent with $\bm{\alpha \approx -1.5}$ \citep{chiang}. We calculate the individual MMEN slope for each system in our sample and find that the median value is $\bm{\alpha_{med} = -1.91}$, which falls between the two aformentioned values but holds greater proximity to the ideally-uniform case. The fact that $\bm{\alpha_{med}}$ is shallower than but still in close to the $\bm{\alpha = -2}$ expectation indicates the dominant trend of uniformity while simultaneously suggesting that inner planets are generally less massive than their outer companions, which is consistent with the findings of planetary mass-ordering presented in \citet{millholland}.

$\log(\Sigma_{0})$ quantifies the solid surface density of disk material at 1 AU, but since nearly all of our systems are fully contained within this distance and disk morphology has been shown to exhibit little variation between such systems \citep{chiang}, we may use the characteristic solid surface density $\log(\Sigma_{0})$ as an indicator of the total mass budget of a given system. We calculate the mass Gini $\mathcal{G}_{M}$ for each system and plot in Figure \ref{fig:5} the relationship between $\log(\Sigma_{0})$ and $\mathcal{G}_{M}$, where we observe a distinctly positive correlation ($r = 0.25$) that suggests a general increase in system mass diversity with increasing characteristic solid surface density. The trend displayed by these systems can be most succinctly be interpreted as the following: \textit{close-in, compact multiple-planet systems with greater non-uniformity in their constituent planetary masses may have formed from protoplanetary disks with a greater total mass budget}, in accord with the formalism presented by \cite{adams}. 


\section{Acknowledgments} \label{sec:acknowledgments}

We thank Sarah Millholland, Fred Adams, Gregory Laughlin, Josh Winn, Fei Dai, and the anonymous referee for their detailed feedback and suggestions for this manuscript. We thank Fred Adams for his invaluable insight into the energy optimization formalism provided by his lightning talk and subsequent personal discussion at the University of Michigan's 2021 Great Lakes Exoplanet Area Meeting (GLEAM). We thank Brandon Radzom for his commentary and support throughout various stages of this work.

\bibliographystyle{aasjournal}
\bibliography{main}



\end{document}